\documentclass[10pt,twocolumn,english,aps,prd,nofootinbib, eqsecnum]{revtex4}

\usepackage[utf8]{inputenc}
\usepackage{graphicx}
\usepackage{amsmath}
\usepackage{amssymb}
\usepackage{amsfonts}
\usepackage{lscape}
\usepackage{hyperref}
\usepackage{url}
\usepackage{epsfig}
\usepackage{color}
\usepackage{bm}
\usepackage{tabularx}
\usepackage{braket}
\usepackage{xcolor}
\newcommand{\be}{\begin{equation}}
\newcommand{\ee}{\end{equation}}
\newcommand{\ba}{\begin{eqnarray}}
\newcommand{\ea}{\end{eqnarray}}

\renewcommand{\[}{\begin{equation}}
\renewcommand{\]}{\end{equation}}

\begin{document}

\thispagestyle{empty}

\title{Covariant formulation of non-equilibrium thermodynamics in General Relativity}

\author{Lloren\c{c} Espinosa-Portal\'es}\email[]{llorenc.espinosa@uam.es}
\author{Juan Garc\'ia-Bellido}\email[]{juan.garciabellido@uam.es}

\affiliation{Instituto de F\'isica Te\'orica UAM-CSIC, Universidad Auton\'oma de Madrid,
Cantoblanco, 28049 Madrid, Spain}

\date{\today}

\begin{abstract}
We construct a generally-covariant formulation of non-equilibrium thermodynamics in General Relativity. We find covariant entropic forces arising from gradients of the entropy density, and a corresponding non-conservation of the energy momentum tensor in terms of these forces. We also provide a Hamiltonian formulation of General Relativity in the context of non-equilibrium phenomena and write the Raychaudhuri equations for a congruence of geodesics. We find that a fluid satisfying the strong energy condition could avoid collapse for a positive and sufficiently large entropic-force contribution. We then study the forces arising from the internal production of ``bulk" entropy of hydrodynamical matter, as well as from the entropy gradients in the boundary terms of the action, like those associated with black hole horizons. Finally, we apply the covariant formulation of non-equilibrium thermodynamics to the expanding universe and obtain the modified Friedmann equations, with an extra term corresponding to an entropic force satisfying the second law of thermodynamics.
\end{abstract}
\maketitle

\section{Introduction}

Einstein’s theory of General Relativity (GR) was formulated more than a century ago and it is still one of the most successful theories in the history of Physics. Its description of the gravitational interaction as a manifestation of geometry has passed all observational tests so far~\cite{Berti:2015itd}. Together with the Standard Model (SM) of Particle Physics, these two theories provide a complete description of Nature at its fundamental level, at least to the extent to which our experiments and observations can reach~\cite{Zyla:2020zbs}. Only two observed phenomena lack a fully satisfactory explanation: the existence of dark matter and the accelerated expansion of the universe. However, they are by themselves inconclusive when arguing for an extension of the GR+SM description of fundamental physics.

It is true, however, that the existence of space-time singularities in GR challenges the validity of the theory around them, where one also expects curvature to be extremely high~\cite{Wald:1984rg}. Furthermore, GR and the SM may provide inconsistent descriptions of Nature. On the one hand, GR is a classical theory that lacks a UV-complete quantum counterpart~\cite{Wald:1984rg}, whereas the SM is built on the framework of Quantum Field Theory (QFT). One can indeed consistently define a QFT on a geometric background given by GR and even compute some effects that the quantum and the classical field have on each other~\cite{Mukhanov:2007zz}, but this is still not a full quantum description. On the other hand, the coupling of SM particles, in particular the Higgs boson, to potential quantum gravitational degrees of freedom (d.o.f.) at energies around the Planck scale is related to the so-called hierarchy problem~\cite{ArkaniHamed:1998rs}.

Proposals for a UV-complete theory of quantum gravity have been extensively explored for decades. Interestingly, GR itself provides already some insight to the need of this quantum description. The work of 
Hawking~\cite{Hawking:1974sw} and 
Bekenstein~\cite{Bekenstein:1973ur} introduced the notion of temperature and entropy of a black hole, leading to the formulation of black hole thermodynamics~\cite{Bardeen:1973gs}. This points towards the existence of unknown microphysical quantum d.o.f., being the geometric description of gravity an emergent macrophysical phenomenon. The link between gravity and thermodynamics has only grown ever since. It has been argued that it constitutes the first piece of the connection between classical and quantum gravity~\cite{Padmanabhan:2009vy}.  The discovery of the area law of entanglement entropy~\cite{Bombelli:1986rw, Srednicki:1993im} particularly supports this idea.

Motivated by the relevance of thermodynamics in gravity, we argue for the need of a proper understanding of the interplay between GR and non-equilibrium thermodynamics. GR, like other physical theories that can be deduced from the stationary action principle, is a time-reversible theory. It is true that the dynamics of horizons has irreversible features, as dictated for instance by the already mentioned black hole thermodynamics, in particular the second law~\cite{Hawking:1971vc}. Still, irreversible phenomena are not included into GR in a complete and systematic way. It is the purpose of the work presented in this paper to provide such an inclusion, i.e. a covariant formulation of non-equilibrium thermodynamics in GR. Our results show that non-equilibrium phenomena, either in the matter content or space-time itself, lead to a back-reaction on the gravitational field equations with potential observational consequences.

This paper is organized as follows. In section II we review existing work on the variational formulation of non-equilibrium thermodynamics. In section III we apply this concept to gravity and show how it fits with both the Lagrangian and Hamiltonian formulation of GR. In section IV we argue that temperature and entropy are naturally included in the matter or gravitational Lagrangian. In section V we look for applications of our results and obtain the non-equilibrium Friedmann and Raychaudury equations. We finish with our conclusions in section VI.

\section{Variational formulation of non-equilibrium thermodynamics}

We begin our work by reviewing the variational formulation for non-equilibrium thermodynamical systems, which was developed by Gay-Balmaz and Yoshimura in ~\cite{GAYBALMAZ2017169, GAYBALMAZ2017194}. Such formulation is the synthesis of two physical frameworks. On the one hand, one has the laws of thermodynamics in the axiomatic formulation of St\"uckelberg:

\begin{itemize}
    \item \emph{First law} or energy conservation. For every thermodynamic system there is an extensive scalar quantity $E$ called \emph{energy}, which can only change due to interactions with the environment:
    \begin{equation}
        \frac{dE}{dt} = P^{ext}(t)\,.
    \end{equation}
    \item \emph{Second law} or positive entropy production. For every thermodynamic system there is an extensive scalar quantity $S$ called \emph{entropy}, which is a monotonically increasing function of time:
    \begin{equation}
        \frac{dS}{dt} = I(t) \ge 0\,,
    \end{equation}
    where the equality holds only once the system is in equilibrium.
\end{itemize}

On the other hand, the dynamics of a mechanical system is dictated by the stationary action principle:

\begin{itemize}
    \item \emph{Stationary action principle}. For a dynamical system with configuration manifold $Q$ and Lagrangian function $L(q,\dot{q}): TQ \rightarrow \mathbb{R}$, the physical curve defined on an interval $t \in [t_1, t_2]$, satisfies the variational condition:
    \begin{equation}
    \delta \int_{t_1}^{t_2} dt L(q, \dot{q}) = 0  \,,
    \end{equation}
    which delivers the well-known Euler-Lagrange equations:
    \begin{equation}
        \frac{d}{dt}\frac{\partial L}{\partial \dot{q}} - \frac{\partial L}{\partial q} = 0\,.
    \end{equation}
\end{itemize}

In order to merge these principles, let us now consider a mechanical system whose Lagrangian function depends on the entropy $S$ as well. The stationary action principle dictates that the physical curve $(q(t), S(t))$ on $Q \times \mathbb{R}$, defined on an interval $t \in [t_1, t_2]$, satisfies the variational condition: 
\begin{equation}
\label{eq:var_prin}
    \delta \int_{t_1}^{t_2} dt L(q, \dot{q}, S) = 0  \,.
\end{equation}
Since the thermodynamical system is out of equilibrium, it must be supplemented with the laws of thermodynamics. Energy conservation is related to the symmetry of the Lagrangian under time translations and so is already encoded in the variational formulation. On the other hand, the second law requires the introduction of a friction or entropic force $F: TQ \times \mathbb{R} \rightarrow T^* Q$ and is implemented by a variational constraint:
\begin{equation}
\label{eq:var_cons}
    \frac{\partial L}{\partial S} (q, \dot{q}, S) \delta S = \left<F(q, \dot{q}, S), \delta q \right>\, ,
\end{equation}
where $\left< \cdot , \cdot \right>$ denotes the scalar product. This variational constraint comes also with a \textit{phenomenological} constraint:
\begin{equation}
\label{eq:phen_cons}
    \frac{\partial L}{\partial S} (q, \dot{q}, S) \dot{S} = \left<F (q, \dot{q}, S), \dot{q} \right> \,.
\end{equation}
The curve $(q(t), S(t))$ that satisfies all three conditions is given by:
\begin{equation}
\begin{aligned}
    &\frac{d}{dt} \frac{\partial L}{\partial \dot{q}} - \frac{\partial L}{\partial q} = F (q, \dot{q}, S)\\
    &\frac{\partial L}{\partial S} \dot{S} = \left<F (q, \dot{q}, S), \dot{q} \right> \,.
\end{aligned}
\end{equation}
Hence, once the effect of non-equilibrium thermodynamics is enforced by the second law, one obtains the Euler-Lagrangian equation with an additional force of entropic origin.
This variational formulation of non-equilibrium thermodynamics applies as it is to isolated systems, i.e. thermodynamic systems that do not exchange energy (heat and work) nor matter with its environment. In order to consider a closed thermodynamic system, i.e. one that exchanges energy but not matter with its environment, one must include the effect of external work in eq. \ref{eq:var_prin} and external heat supply in eq. \ref{eq:phen_cons}. The generalization to open systems, i.e. that allow both energy and matter exchange, is developed in \cite{Gay-Balmaz_2018}.
As we will see later, one often finds that the \emph{temperature} of the thermodynamic system can be introduced as:
\begin{equation}
    \frac{\partial L}{\partial S} = - T \,,
\end{equation}
although it is not necessary to do so in the general case.

The thermodynamical nature of the system implies the explicit or implicit coarse-grain of some d.o.f. Its detailed microphysics may be unknown, but its macrophysics has an effect on the dynamics of other physical variables whose microphysics is known. This effect is encoded by the second law of thermodynamics, which restricts the configuration space and delivers a modified equation of motion, and it allows us to study the coupling between these d.o.f.

The full variational implementation of the continuum case is a bit more involved, should the entropy be allowed to have a spatial dependence \cite{GAYBALMAZ2017194}. The reason is that friction causes internal entropy production, but entropy can also increase or decrease locally due to entropy fluxes. For the sake of clarity and when required, we will instead follow a short-cut and introduce the required additional equations from physical considerations.

\section{Non-equilibrium dynamics in General Relativity}

After reviewing the variational formulation of non-equilibrium thermodynamics we are ready to apply this same formalism to General Relativity. The coupling of the gravitational field to coarse-grained physical d.o.f. delivers an effective modification of Einstein field equations. We will first show this by supplementing the Einstein-Hilbert action with the constraints given by the second law of thermodynamics. Then, we will check that it is also consistent with the Hamiltonian formulation of General Relativity. We will also provide a physical insight onto the effects of this effective modification of the gravitational dynamics by inspecting the Raychauduri equation.

\subsection{Lagrangian formulation}

The variational formalism can be applied directly to the Einstein-Hilbert action without any particular assumption on the metric. However, it requires the introduction of a foliation of the space-time manifold. This cannot be avoided: the second law of thermodynamics is linked to the existence of the arrow of time. Nevertheless, the Einstein field equations keep its general covariance, as we will check shortly.

Let us build our action as the sum of the Einstein-Hilbert action of General Relativity plus a matter term:
\begin{equation}
\label{eq:grm}
    \frac{1}{2\kappa} \int d^4x \sqrt{-g} R + \int d^4x \mathcal{L}_{m}(g_{\mu \nu}, S)\,,
\end{equation}
where the coupling is $\kappa =8\pi G$ and we allow the matter Lagrangian $\mathcal{L}_m$, which is a tensor density, to have a dependency on the entropy $S$. It may as well depend on additional fields that describe the matter content. In this setup, the stationary action principle takes the form $\delta (\ref{eq:grm}) = 0$. That is:
\begin{equation}
\begin{aligned}
    \int d^4x \left(\frac{1}{2\kappa} \frac{\delta (\sqrt{-g}R)}{\delta g^{\mu \nu}} + \frac{ \delta \mathcal{L}_m}{\delta g^{\mu \nu}}\right)\delta g^{\mu \nu} \\
    + \int d^4x  \frac{\partial \mathcal{L}_m}{\partial S} \delta S = 0\,,
\end{aligned}
\end{equation}
which is now supplemented with the variational constraint given by the second law of thermodynamics:
\begin{equation}
\begin{aligned}
    & \frac{\partial L_{m}}{\partial S} \delta S = \frac{1}{2} F_{\mu \nu} \delta g^{\mu \nu}\,,
\end{aligned}
\end{equation}
where $F_{\mu \nu}$ is the tensorial friction or entropic force. Analogously to the Lagrangian density one can define the friction density as
\begin{equation}
    F_{\mu \nu} =
    \int d^3x \sqrt{-g} f_{\mu \nu}\,.
\end{equation}
The constrained stationary action principle gives the non-equilibrium Einstein field equations:
\begin{equation}\label{ModEinsteinEqs}
    R_{\mu \nu} - \frac{1}{2} R g_{\mu \nu} = \kappa \left( T_{\mu \nu} -  f_{\mu \nu} \right)\,,
\end{equation}
which includes the usual geometric and matter terms plus an entropic one. This equation is one of the main results of our work and shows how non-equilibrium thermodynamics is very relevant in gravitation.

Note that the Bianchi identities, a reflection of the general covariance of the theory, allow the covariant non-conservation of the energy-momentum tensor
\be\label{nonconservation}
D^\mu T_{\mu\nu} = D^\mu f_{\mu\nu}\,.
\ee
%

One can include a dependence of the entropy on the spatial position by introducing the entropy density $s(\vec{x},t)$ and rewriting the variational constraint as:
\begin{equation}
    \frac{\partial  \mathcal{L}_m }{\partial s} \delta s = \frac{1}{2} \sqrt{-g} f_{\mu \nu} \delta g^{\mu \nu}\,,
\end{equation}
provided that there is no dependence of the Lagrangian on the partial derivatives $\partial_\mu s$.

For now, we have shown that the variational constraint is enough to obtain the non-equilibrium Einstein field equations. We will deal with the phenomenological constraint in the next subsection, once a foliation of space-time is explicitly introduced in the context of the ADM formalism. The phenomenological constraint will allow us to obtain an implicit expression for the force $f_{\mu \nu}.$

The fully rigorous implementation of the variational constraint becomes a bit more subtle once this spatial dependence is introduced. In practice, the function $s$ is not the entropy itself, but rather a function that plays a role in the entropy balance equation. In particular $\delta s$ is the entropy density variation due to internal processes and not to entropy fluxes. We leave the discussion of this equation to the next section, as it also requires the introduction of a foliation. 

\subsection{Hamiltonian formulation}

The variational formalism delivers a modification of Einstein field equations due to the appearence of a force of entropic origin. We will make the effect of this entropic force more concrete by employing the Hamiltonian formulation of General Relativity or Arnowitt-Deser-Misner (ADM) formalism. That is, we perform a (3+1)-splitting of space-time, a foliation that parametrizes the 4-dimensional metric $g_{\mu \nu}$ by means of a 3-dimensional metric $h_{ij}$ and the lapse and shift functions $N$ and $N^i$. Space-time dynamics is treated as the evolution of space-like hypersurfaces $\Sigma_t$, parametrized by some parameter $t$, which is usually taken to be the time coordinate. For more details, see e.g. \cite{Wald:1984rg}. In this formalism, an arbitrary metric takes the form:
\begin{equation}
    ds^2 = - (Ndt)^2 + h_{ij} (dx^i + N^i dt) (dx^j + N^j dt)\,.
\end{equation}
We will denote as $\Sigma$ the 3-dimensional hypersurface and $n$ its normal vector:
\begin{equation}
    n_\alpha = (-N, 0, 0, 0)\,,
\end{equation}
which is a unit vector, i.e. $n_\alpha n^\alpha = -1$. Space-time indices are lowered and raised as usual by $g_{\mu \nu}$. Spatial indices, however, are lowered and raised by $h_{ij}$, which furthermore satisfies $h_{ij} h^{jk} = \delta_i^k$.

Equivalently, one can write the splitting of the metric as:
\begin{equation}
    h_{\mu \nu} = g_{\mu \nu} + n_{\mu} n_{\nu}\,,
\end{equation}
so that it is clear that $h_{\mu \nu}$ is purely tangential to the hypersurface. Then its spatial part $h_{ij}$ is equal to the pull-back of the 4-dimensional metric $g_{\mu \nu}$ onto $\Sigma$ and is a legitimate 3-dimensional metric.

The Einstein-Hilbert action for this parametrization of the metric is given by the following gravitational Lagrangian:
\begin{equation}
    \mathcal{L}_G = \sqrt{-g} R =N \sqrt{h} \left(^{(3)}R + K_{ij}K^{ij} - K^2 \right)\,,
\end{equation}
where $K_{ij}$ is the extrinsic curvature of the 3-hypersurface $\Sigma$ and is given by the Lie derivative along the normal vector $n$:
\begin{equation}
    K_{ij} = \frac{1}{2} \pounds_n h_{ij} = \frac{1}{2N} \left(\partial_0 h_{ij} - \nabla_i N_j - \nabla_j N_i \right)\,.
\end{equation}
where $\nabla$ denotes the covariant derivative on $\Sigma$ with respect to the 3-metric $h_{ij}$. Its trace and traceless part are:
\begin{equation}
\begin{aligned}
    &K = h^{ij}K_{ij} = \frac{1}{N} \left(\partial_0 \ln \sqrt{h} - \nabla_i N^i\right)\\
    &\bar{K}_{ij} = K_{ij} - \frac{1}{3} K h_{ij}\,.
\end{aligned}
\end{equation}
Unlike the intrinsic curvature, described by the Riemann tensor $R^{\rho}_{\mu \nu \lambda}$ and its contractions, the extrinsic curvature is a quantity that depends on the embedding of a surface in a larger manifold.

The extrinsic curvature can be a complicated function of the parameters. Therefore, it is convenient to shift to the Hamiltonian formulation of the stationary-action principle. Note that the only quantity whose time derivative appears in the gravitational Lagrangian is the 3-spatial metric $h_{ij}$ and, thus, it is the only dynamical or propagating d.o.f. Correspondingly, one defines its conjugate momentum as:
\begin{equation}
    \Pi^{ij} =  \frac{\partial \mathcal{L}_G}{\partial \dot{h}_{ij}} =  \sqrt{h}\left( K^{ij} - K h^{ij} \right)\,.
\end{equation}
With this, the gravitational Lagrangian can be rewritten as:
\begin{equation}
\begin{aligned}
\mathcal{L}_G & =  N\sqrt{h} ^{(3)} R - \frac{N}{\sqrt{h}}\left(\Pi_{ij} \Pi^{ij} -\frac{1}{2} \Pi^2 \right) - 2 \Pi^{ij} \nabla_i N_j \\
   & = \Pi^{ij} \dot{h}_{ij} -  N \mathcal{H} - N_i \mathcal{H}^i - 2 \nabla_i \left(\Pi^{ij} N_j \right)\,,
\end{aligned}
\end{equation}
where $\Pi = h_{ij} \Pi^{ij}$ and we introduced the functions:
\begin{equation}
\begin{aligned}
    & \mathcal{H} = - \sqrt{h} ^{(3)} R + \frac{1}{\sqrt{h}} \left( \Pi_{ij}\Pi^{ij} - \frac{1}{2}\Pi^2 \right)\\
    & \mathcal{H}^{i} = -2 \nabla_j \left(h^{-1/2} \Pi^{ij}\right)\,.
\end{aligned}
\end{equation}
Since $N$ and $N_i$ are not dynamical variables, they merely enter the gravitational Lagrangian as Lagrange multipliers. One defines the gravitational Hamiltonian as:
\begin{equation}
\begin{aligned}
    \mathcal{H}_G & = \Pi^{ij} \dot{h}_{ij} - \mathcal{L}_G\\
    & = N \mathcal{H} + N_i \mathcal{H}^i + \nabla_i \left(\Pi^{ij} N_j \right)\,,
\end{aligned}
\end{equation}
with the Hamiltonian and momentum constraints:
\begin{equation}
\begin{aligned}
    &\frac{\delta \mathcal{H}_G}{\delta N} =  \mathcal{H} = 0\\ &\frac{\delta \mathcal{H}_G}{\delta N_i} =  \mathcal{H}^i = 0\,.
\end{aligned}
\end{equation}
The evolution equations are obtained upon taking variations of the Hamiltonian, but these need to be modified once the second law of thermodynamics is enforced and entropic forces come into play. The first Hamilton equation is:
\begin{equation}
\begin{aligned}
    \frac{\delta \mathcal{H}_G}{\delta \Pi^{ij}} & = \dot{h}_{ij} - \frac{\partial \dot{h}_{kl}}{\partial \Pi^{ij}}\Pi^{kl} - \frac{\partial \mathcal{L}_G}{\partial \dot{h}_{kl}}\frac{\partial \dot{h}_{kl}}{\partial \Pi^{ij}}\\
    & = \dot{h}_{ij}\,.
\end{aligned}
\end{equation}
This equation is true regardless of the constraint imposed by the second law of thermodynamics. The second Hamilton equation will carry the effect of non-equilibrium thermodynamics. Let us compute it starting from the derivative:
\begin{equation}
\begin{aligned}
    \frac{\partial{\mathcal{H}_G}}{\partial h_{ij}} & = \Pi^{kl} \frac{\partial \dot{h}_{kl}}{\partial h_{ij}} - \frac{\partial \mathcal{L}_G}{\partial \dot{h}_{kl}}\frac{\partial  \dot{h}_{kl}}{\partial h_{ij}} - \frac{\partial \mathcal{L}_G}{\partial h_{ij}}\\
    & = - \frac{\partial \mathcal{L}_G}{\partial h_{ij}} = - \frac{\partial \mathcal{L}_G}{\partial h^{kl}} \frac{\partial h^{kl}}{\partial h_{ij}}\,.
\end{aligned}
\end{equation}
One usually applies the field equation in order to obtain the second Hamilton equation (see e.g. \cite{jose_saletan_1998}). Here, we will do the same but taking into account the constraints imposed by the second law of thermodynamcs. The 3+1 splitting of the space-time manifold allows us to identify the 3-metric $h_{ij}$ as the dynamical d.o.f. Then we argue that the phenomenological constraint should involve only those and, therefore, relates their dynamical evolution to changes in the entropy density:
\begin{equation}
    \frac{\partial \mathcal{L}}{\partial s} \pounds_n s = \frac{1}{2} N\sqrt{h} \tilde{f}_{ij} \pounds_n h^{ij}\,.
\end{equation}
The Lie derivative $\pounds_n$ along the normal vector $n$ serves here as a generalization of the time derivative. Note that the same equation for the phenomenological constraint is obtained if one characterizes the evolution of the hypersurfaces by the flow along the vector $m = N n$, which may be preferred as it satisfies $g (\partial_t, m) = 1$. 
The variational constraint should only involve dynamical d.o.f. as well:
\begin{equation}
\begin{aligned}
    \frac{\partial \mathcal{L}}{\partial s} \delta s = \frac{1}{2} N\sqrt{h} \tilde{f}_{ij} \delta h^{ij}\,.
\end{aligned}
\end{equation}
The tensor $\tilde{f}_{ij}$ should now be understood as the pull-back of the projection of $f_{\mu\nu}$ on $\Sigma$, i.e.
\begin{equation}
    \tilde{f}_{ij} = h^\mu_i h^\nu_j f_{\mu \nu}   
\end{equation}
We argue that this is the only non-vanishing part of $f_{\mu\nu}$. Our claim is supported by the fact that
\begin{equation}
    \pounds_n (n ^{\mu} n ^{\nu}) = 0
\end{equation}
The tensor $\tilde{f}_{ij}$ will have contributions from its trace and trace-less component according to the tensor decomposition:
\begin{equation}
    \tilde{f}_{ij} = \frac{1}{3} \tilde{f} h_{ij} + \nabla_{(i} \tilde{f}_{j)} + \nabla_i \nabla_j \bar{f} + f^{TT}_{ij}\,,
\end{equation}
where
\begin{equation}
\label{eq:ten_deco}
    \tilde{f} = \tilde{f}_{ij} h^{ij}\quad
    \nabla^i \tilde{f}_i = 0\quad
    \nabla^i \nabla_i \bar{f} = 0 \quad \nabla^{i}f^{TT}_{ij} = h^{ij} f^{TT}_{ij} = 0\,.
\end{equation}
We point out now that the interpretation of $s(t, \vec{x})$ as the entropy density is subtle. Instead, it is a function that satisfies the following entropy balance equation
\begin{equation}
    \pounds_n s = \pounds_n s^{tot} - \nabla_i j_s^i\,,
\end{equation}
where $\pounds_n s^{tot}$ is the total entropy production and $j_s$ is the entropy flux in the hypersurface. Hence, $\pounds_n s$ is interpreted as the internal entropy production. It is important to bear this balance in mind when applying the phenomenological constraint to actual physical scenarios. We refer the interested reader to the appendix and to the original work on the variational formulation of non-equilibrium thermodynamics in the continuum \cite{GAYBALMAZ2017194} for the description of a fully variational formulation and its consistency with the shortcut version presented here.

Once we have properly taken into account the constraints given by the second law of thermodynamics, we are ready to obtain the field equation for the 3-metric as
\begin{equation}
    \frac{\delta \mathcal{L}_G}{\delta h^{ij}} = \frac{\partial \mathcal{L}_G}{\partial h^{ij}} - \partial_{\mu} \frac{\partial \mathcal{L}_G}{\partial \partial_{\mu} h^{ij}} = - \frac{1}{2} N \sqrt{h} \tilde{f}_{ij}\,.
\end{equation}
One can then recast it as:
\begin{equation}
    \frac{\partial \mathcal{L}_G}{\partial h^{kl}}\frac{\partial h^{kl}}{\partial h_{ij}} = \left( \partial_{\mu} \frac{\partial \mathcal{L}_G}{\partial \partial_{\mu} h^{kl}} - \frac{1}{2} N \sqrt{h} \tilde{f}_{kl} \right) \frac{\partial h^{kl}}{\partial h_{ij}}
\end{equation}
and rewrite the derivative:
\begin{equation}
\begin{aligned}
    \frac{\partial \mathcal{H}_G}{\partial h_{ij}} =& - \left( \partial_{\mu} \frac{\partial \mathcal{L}_G}{\partial \partial_{\mu} h^{kl}} - \frac{1}{2} N \sqrt{h}\tilde{f}_{kl} \right) \frac{\partial h^{kl}}{\partial h_{ij}}\,.
\end{aligned}
\end{equation}
Now, by using the relations:
\begin{equation}
\begin{aligned}
    &\frac{\partial h_{ij}}{\partial h^{kl}} = -\frac{1}{2} \left(h_{i k} h_{j l} + h_{i l} h_{j k}\right)\\
    &\frac{\partial \partial_\mu h_{i j}}{\partial \partial_\nu h^{k l}} = -\frac{1}{2} \delta^{\nu}_\mu \left(h_{i k} h_{j l} + h_{i l} h_{j k}\right)
\end{aligned}
\end{equation}
we get
\begin{equation}
\begin{aligned}
	\frac{\partial \mathcal{H}_G}{\partial h_{ij}}
	& = - \dot{\Pi}^{ij} + \partial_k \frac{\partial \mathcal{H}_G}{\partial \partial_k h_{ij}} - \frac{1}{2} N \sqrt{h} \tilde{f}^{ij}
\end{aligned}
\end{equation}
and introducing the functional derivative
\begin{equation}
    \frac{\delta \mathcal{H}_G}{\delta h_{ij}} = \frac{\partial \mathcal{H}_G}{\partial h_{ij}} - \partial_{\mu} \frac{\partial \mathcal{H}_G}{\partial \partial_\mu h_{ij}}
\end{equation}
we obtain the second Hamilton equation:
\begin{equation}
    \frac{\delta \mathcal{H}_G}{\delta h_{ij}} = - \dot{\Pi}^{ij} - \frac{1}{2} N\sqrt{h}\tilde{f}^{ij}\,.
\end{equation}
This equation is modified by the existence of an entropic force in consistency with the thermodynamical constraint. Physically, it is this equation that describes the dynamical evolution, for it implicitly contains a second time derivative of the field. 

Matter can be included in the Lagrangian density as
\begin{equation}
    \mathcal{L} = \mathcal{L}_G(h_{ij}, \dot{h}_{ij}) + 2\kappa\,\mathcal{L}_m(h_{ij}, S)\,,
\end{equation}
where we introduced the gravitational coupling in the matter Lagrangian for convenience. Then the Hamilton equations become
\begin{equation}
\begin{aligned}
    &\frac{\delta \mathcal{H}_G}{\delta \Pi^{ij}} = \dot{h}_{ij}\\
    &\frac{\delta \mathcal{H}_G}{\delta h_{ij}} = - \dot{\Pi}^{ij} - 2\kappa\, \frac{\delta \mathcal{L}_m}{\delta h_{ij}} - \kappa\, N \sqrt{h}\tilde{f}^{ij}\,,
\end{aligned}
\end{equation}
where the entropic term carries the coupling $\kappa$ as well if it comes from the matter Lagrangian. The Hamiltonian and momentum constraints are likewise modified by the introduction of matter:
\begin{equation}
\begin{aligned}
    \frac{\delta \mathcal{H}_G}{\delta N} = \mathcal{H} = 2\kappa \frac{\partial{\mathcal{L}_m}}{\partial N}\\
    \frac{\delta \mathcal{H}_G}{\delta N_i} = \mathcal{H}^i = 2\kappa \frac{\partial{\mathcal{L}_m}}{\partial N_i}\,.
\end{aligned}
\end{equation}
The variational formalism for non-equilibrium thermodynamics developed in \cite{GAYBALMAZ2017169, GAYBALMAZ2017194} fits nicely in both the Lagrangian and Hamiltonian formulation of General Relativity. This means that one can naturally consider effects of non-equilibrium thermodynamics in General Relativity and obtain analytical or numerical solutions to the equations of motion.

\subsection{The Raychauduri equations}

The appearance of an entropic term in Einstein's field equations can have dynamical effects that may look as a violation of the energy conditions. We wil look now into this possibility by studying a congruence of worldlines in an arbitrary space-time. These need not be geodesics and have tangent vector $n$. The congruence is then characterized by the tensor:
\begin{equation}
    \Theta_{\mu \nu} = D_\nu n_\mu = \frac{1}{3} \Theta h_{\mu \nu} + \sigma_{\mu \nu} + \omega_{\mu \nu} - a_{\mu}n_\nu \,
\end{equation}
where $\theta$ is the expansion rate of the congruence, $\sigma_{\mu \nu}$ is its shear or symmetric trace-less part and $\omega_{\mu \nu}$ is its vorticity or antisymmetric part. If the worldline is not a geodesic, then the congruence suffers an acceleration given by:
\begin{equation}
    a_{\mu} = n^{\nu} D_{\nu} n_{\mu}
\end{equation}
One can compute the Lie derivative of the expansion of the congruence along its tangent vector and find the Raychauduri equation \cite{Wald:1984rg}:
\begin{equation}
    \pounds_n \Theta = - \frac{1}{3} \Theta^2 - \sigma_{\mu \nu} \sigma^{\mu \nu} + \omega_{\mu \nu} \omega^{\mu \nu} - R_{\mu \nu} n^{\mu} n^{\nu} + D_{\mu} a^{\mu}\,.
\end{equation}
Let us perform the standard analysis of the sign of this equation. It is clear that $\sigma_{\mu \nu} \sigma^{\mu \nu} > 0$ and $\Theta^2 >0$. On the other hand, if the congruence is chosen to be orthogonal to the spatial hypersurfaces, as we have been considering, then the vorticity vanishes $\omega_{\mu \nu} = 0$. Lastly, it is left to consider the term $R_{\mu \nu} n^{\nu} n^{\nu}$, which we can rewrite with the help of the field equations:
\begin{equation}
    R_{\mu \nu} n^{\mu} n^{\nu} = 8\pi G \left( T_{\mu \nu}n^{\mu}n^{\nu}  + \frac{1}{2} T - f_{\mu \nu}n^{\mu}n^{\nu}  - \frac{1}{2} f \right)
\end{equation}
If the strong energy condition is satisfied, then:
\begin{equation}
    T_{\mu \nu} n^{\mu} n^{\nu} \ge - \frac{1}{2} T
\end{equation}
and, in the absence of intrinsic acceleration, $a_\mu=0$, we can establish the bound:
\begin{equation}
    \pounds_n \Theta + \frac{1}{3} \Theta^2 \le 8\pi G \left( f_{\mu \nu}n^\mu n^\nu + \frac{1}{2} f \right)
\end{equation}
For a vanishing entropic force $f_{\mu \nu} = 0$, this means that an expanding congruence cannot indefinitely sustain its divergence and will eventually recollapse. On the contrary, a positive and sufficiently large entropic contribution can avoid such recollapse. This may become relevant for an expanding universe, but also to generic gravitational collapse and the singularity theorems \cite{Penrose:1964wq, Penrose:1969pc, Hawking:1969sw}.

The shear $\sigma_{\mu \nu}$ is also affected by the inclusion of an entropic force. Its evolution equation is given by
\begin{equation}
\begin{aligned}
    \pounds_n \sigma_{\mu \nu} = & - \frac{2}{3} \theta\,\sigma_{\mu \nu} - \sigma_{\mu \lambda} \sigma^{\lambda}_{\ \nu} - \omega_{\mu \lambda} \omega^{\lambda}_{\ \nu} + C_{\lambda\mu\rho\nu} n^\lambda n^\rho  \\
    & +\frac{1}{3} h_{\mu \nu} (\sigma_{\lambda \rho} \sigma^{\lambda \rho} - \omega_{\lambda \rho} \omega^{\lambda \rho}) + \frac{1}{2}\hat{R}_{\mu \nu}\,,
\end{aligned}
\end{equation}
where $C_{\lambda\mu\rho\nu}$ is the Weyl tensor and $\hat{R}_{\mu \nu}$ is the spatial, trace-free part of the Ricci tensor:
\begin{equation}
    \hat{R}_{\mu \nu} = h_{\mu \lambda} h_{\nu \rho} R^{\lambda \rho} - \frac{1}{3} h_{\mu \nu} h_{\lambda \rho} R^{\lambda \rho}
\end{equation}
This explicit dependence on the Ricci tensor allows us to directly include the effect of the entropic force and establish a bound. Indeed, using the modified Einstein field equation we get
\begin{equation}
    \hat{R}_{\mu \nu} = 8\pi G \left( \hat{T}_{\mu \nu} - \hat{f}_{\mu \nu} \right)\,,
\end{equation}
where $\hat{f}_{\mu \nu}$ and $\hat{T}_{\mu \nu}$ are the analogously defined spatial, trace-free part of the friction and stress-energy tensors. Then we can rewrite the evolution equation for the shear as:
\begin{equation}
\begin{aligned}
    \pounds_n \sigma_{\mu \nu} = & - \frac{2}{3} \theta\,\sigma_{\mu \nu} - \sigma_{\mu \lambda} \sigma^{\lambda}_{\ \nu} - \omega_{\mu \lambda} \omega^{\lambda}_{\ \nu} + C_{\lambda\mu\rho\nu} n^\lambda n^\rho  \\
    & +\frac{1}{3} h_{\mu \nu} (\sigma_{\lambda \rho} \sigma^{\lambda \rho} - \omega_{\lambda \rho} \omega^{\lambda \rho}) + 4\pi G \left( \hat{T}_{\mu \nu} - \hat{f}_{\mu \nu} \right)\,,
\end{aligned}
\end{equation}
which means that the entropic force directly sources the shear in a way similar to that of the stress-energy tensor.

These results also apply to a congruence of worldlines which are not normal to the hypersurfaces that define the foliation of the space-time. In that case, the vorticity is non-vanishing and has the evolution equation
\begin{equation}
    \pounds_n \omega_{\mu \nu} = - \frac{2}{3} \theta\,\omega_{\mu \nu} - 2 \sigma^{\lambda}_{[\nu} \omega_{\mu]\lambda}\,, 
\end{equation}
which is not directly sourced by the entropic term. Of course, it is still affected indirectly due to modifications in the metric, the expansion and the shear.

The decomposition of the 2-tensor describing the congruence of geodesics into global expansion, shear and vorticity, which are affected by the entropic forces via the corresponding Raychaudhuri equations, brings to mind the evolution of large scale structures in the cosmic web due to gravitational collapse of initial fluctuations. The growth of structure brings order into an otherwise homogeneous universe, so we expect a corresponding entropy production in the outskirts of large structures like galaxies and clusters of galaxies. According to our formulation, on supergalactic scales, such an entropy production should give rise to a local acceleration, leaving large voids between superclusters, enhancing the contrast induced by the usual gravitational collapse. Moreover, in the formation of the first spiral galaxies there is also an associated entropy production which could give rise to a tiny acceleration, that may explain part of the rotation curves of galaxies, beyond that produced by the dark matter in the halos of galaxies.

\section{Temperature and entropy}

So far we have imposed the second law of thermodynamics by a constraint which contains the derivative $\partial L/ \partial S$. The goal of this section is to understand how this term is often linked to the concept of temperature of a thermodynamical system, not only in a mechanical system but also in General Relativity. In doing so, we will consider two sources of entropy: hydrodynamical matter and gravity itself.

\subsection{Bulk entropy: Hydrodynamical matter}

Let us first consider a mechanical system with Lagrangian given by
\begin{equation}
    L(q,\dot{q}, S) = E_K(q, \dot{q}) - U(q, S)\,,
\end{equation}
where $E_K$ and $U$ are, respectively, the kinetic and internal energy. Notice that only the latter depends on the entropy $S$. One way of obtaining the temperature of this thermodynamic system is by definition:
\begin{equation}
    T = \frac{\partial U}{\partial S} = - \frac{\partial L}{\partial S}\,.
\end{equation}
Then the entropic constraint can also be written as
\begin{equation}
    T \dot{S} = - F \dot{q} > 0\,.
\end{equation}
We can generalize this to a fluid whose Lagrangian is pure internal energy, as it is the case for instance of the cosmic fluid \cite{Mukhanov:1990me}. The matter Lagrangian is then given by
\begin{equation}
    L= \int d^3x \mathcal{L} = - \int d^3x \sqrt{-g} \rho(g_{\mu\nu}, s)\,,
\end{equation}
where $\rho(g_{\mu\nu}, s)$ is the energy density of the fluid. Hence, hydrodynamic matter has a well defined notion of temperature:
\begin{equation}
    T = - \frac{1}{\sqrt{-g}} \frac{\partial \mathcal{L}_m}{\partial s} = - \frac{\partial \rho}{\partial s}\,.
\end{equation}
If the fluid is homogeneous and isotropic this definition is equivalent to
\begin{equation}
    T = - \frac{\partial L}{\partial S}\,.
\end{equation}
The tensor entropic force for a space-time filled with hydrodynamic matter in the ADM formalism is then given implicitly by
\begin{equation}
    F_{ij} \dot{h}^{ij} = - T\dot{S} \le 0\,.
\end{equation}
Since entropy is a monotonically increasing function of time, the second law of thermodynamics constraints the sign of the tensor entropic force.

\subsection{Surface terms: Entropy in the boundary}

One can also wonder about the effect of the entropy associated to space-time itself, in particular to horizons. It can be incorporated in a natural way by extending the Einstein-Hilbert action with a surface term, the Gibbons-Hawking-York (GHY) term of Refs.~\cite{York:1972sj, Gibbons:1976ue}.

Let us consider a space-time manifold $\mathcal{M}$ with metric $g_{\mu \nu}$, which has a horizon hypersurface that we denote by $\mathcal{H}$. This is a submanifold of the whole space-time. By taking  $n^{\mu}$, the normal vector to the hypersurface $\mathcal{H}$, we can define an inherited metric on $\mathcal{H}$:
\begin{equation}
    g_{\mu \nu} = h_{\mu \nu} + n_{\mu} n_{\nu}\,.
\end{equation}
With this, one can define the GHY term as
\begin{equation}
    S_{GHY} = \frac{1}{8\pi G} \int_{\mathcal{H}} d^3 y \sqrt{h} K\,,
\end{equation}
where $K$ is the trace of the extrinsic curvature of the surface. We already considered this quantity when discussing the ADM formalism. Notice, however, that here we are not foliating the entire space-time, but rather considering the properties of a particular hypersurface, the horizon.
From the thermodynamic point of view, the GHY term contributes to the internal energy of the system. Hence, it can be related to the temperature and entropy of the horizon as
\begin{equation}
    S_{GHY} = - \int dt \,N(t)\,T S\,.
\end{equation}
where we have kept the lapse function $N(t)$, to indicate that the variation of the total action with respect to it will generate a Hamiltonian constraint with an entropy term together with the ordinary matter/energy terms. In order to illustrate this, let us now compute the GHY for two horizons of interest: the event horizon of
a Schwarzschild black hole and the horizon of black holes in FLRW universe.

\subsubsection{Schwarzschild black hole}

In order to illustrate this, let us now compute the GHY term for the event horizon of a Schwarzschild black hole of mass $M$. Its space-time is described by the metric:
\begin{equation}
    ds^2 = - \left(1- \frac{2GM}{r} \right) dt^2 + \left(1- \frac{2GM}{r} \right)^{-1} dr^2 + r^2 d\Omega_2^2\,.
\end{equation}
The normal vector to a 2-sphere of radius $r$ around the origin of coordinates is
\begin{equation}
    n = - \sqrt{1 - \frac{2GM}{r}} \partial_r\,.
\end{equation}
With this, the trace of the extrinsic curvature for such a sphere scaled by the metric determinant is
\begin{equation}
    \sqrt{h} K = (3GM - 2r) \sin\theta\,.
\end{equation}
Integrating over the angular coordinates and setting the 2-sphere at the event horizon, i.e. $r = 2GM$, and restoring for a moment $\hbar$ and $c$, the GHY becomes
\begin{equation}
    S_{GHY} = - \frac{1}{2} \int dt Mc^2  = - \int dt T_{BH} S_{BH}\,,
\end{equation}
where $T_{BH}$ is the Hawking temperature and $S_{BH}$ is the Bekenstein entropy of the Schwarzschild black hole:
\begin{equation}
    T_{BH} = \frac{\hbar c^3}{8\pi G M} \quad S_{BH} = \frac{Ac^3}{4G\hbar} = \frac{4\pi G M^2}{\hbar c}\,.
\end{equation}
This favors the interpretation of the GHY term of a horizon as a contribution to the internal energy in the thermodynamic sense. 

\subsubsection{Cosmological black holes}

The natural inclusion of temperature and entropy from surface terms allows is also useful when embedding black holes into an expanding universe. This embedding is not unique \cite{Faraoni:2018xwo}, but one can keep the discussion rather generic by considering the generalized McVittie solution
\begin{equation}
    ds^2 = - N(t)^2 \frac{B(t, \bar{r})^2}{A(t, \bar{r})^2} dt^2 + a(t)^2 A(t, \bar{r})^4 \left( d\bar{r}^2 + \bar{r}^2 d\Omega_2^2 \right)\,,
\end{equation}
where
\begin{equation}
\begin{aligned}
    &A(t,\bar{r}) = 1 + \frac{m(t)}{2 \bar{r}}\\
    &B(t,\bar{r}) = 1 - \frac{m(t)}{2 \bar{r}}\,, 
\end{aligned}
\end{equation}
being $m(t) > 0$ the comoving mass of the black hole, i.e. $m(t) = M(t)/a(t)$, and $N(t)$ the lapse function linked to the residual gauge freedom. Note the use of isotropic coordinates, which are obtained by introducing a new comoving radial coordinate $\bar{r}$, related to the usual areal radius $r$ by
\begin{equation}
    r = a(t) \bar{r}\left( 1 + \frac{m(t)}{2\bar{r}} \right)^2\,.
\end{equation}
For a black hole much smaller than the Hubble scale, its apparent horizon is located at its Schwarzschild radius \cite{Faraoni:2018xwo} and we can assign to it the usual Bekenstein entropy and Hawking temperature. Performing a computation similar to that of the Schwarzschild black hole, we arrive at the following result for the GHY term:
\begin{equation}
    S_{GHY} = - \int dt N(t) T_{BH} S_{BH}\,.
\end{equation}
The growth of black holes comes with an increase in the entropy and an associated entropic force, which may have an impact on the dynamics of the scale factor. Furthermore, if the universe is populated by many black holes, one can compute their average contribution to the stress-energy tensor from these surface terms. Indeed, if one takes now the homogeneous and isotropic flat FLRW metric, which is valid at sufficiently large scales, then the GHY term can be approximated as
\begin{equation}
\begin{aligned}
    S_{GHY} & = - \sum_i \int dt N(t) T_{BH}S_{BH} \simeq\\
    & = - \int d^4 x \sqrt{-g} \,n_{BH} T_{BH}S_{BH} \,,
\end{aligned}
\end{equation}
where $n_{BH}$ is the number density of the black holes. This delivers the following contribution to the stress-energy tensor:
\begin{equation}
    T_{00} =  N(t)^2 \rho_{BH}\,.
\end{equation}
The other components of the stress-energy tensor depend on the accretion onto the black holes. If there is no accretion, as it is the case of the original McVittie metric~\cite{Faraoni:2018xwo}, then the $T_{ij}$ and $T_{0i}$ components vanish and so does the pressure $p$ and we recover the standard interpretation of a collection of black holes as dust. Other accretion conditions may lead to different equations of state.

\subsubsection{Interpretation as a thermodynamic system}

One may interpret the effects of the GHY term as the inclusion of a thermodynamic system. For a localized object like a black hole, its properties are characterised by the Lagrangian:
\begin{equation}
	L = -U\,,
\end{equation}
where $U$ is the internal energy of the system, which we find to be
\begin{equation}
	U = -N TS\,.
\end{equation}
If we ignore the lapse function, linked to the freedom in choosing the time coordinate, this expression is similarly found in usual thermodynamics.

Furthermore, if the thermodynamic system is extended, as in the case of the cosmological black holes, one may interpret this thermodynamic system not as an isolated object but rather as a fluid. In that case the internal thermodynamic energy can be written as a spatial integral of an energy density
\begin{equation}
	U = \int d^3x \sqrt{-g} \rho 	
\end{equation}
and deliver the Lagrangian of a perfect fluid. This fluid satisfies the second law of thermodynamics and may be considered as an effective real fluid after allowing an increase in entropy. In section VI we discuss how the variational formalism includes the theory of real fluids and provides an extension thereof.

\section{Application to Non-equilibrium Cosmology}

Let us now illustrate the potential of our covariant formulation of non-equilibrium thermodynamics in General Relativity by studying how it affects the trajectory of particles. We will consider a particularly relevant example of a space-time, an FLRW universe, and show how the Friedmann equations get modified in this context. These equations directly affect the geodesics followed by inertial observers.

The effect of non-equilibrium thermodynamics in an expanding FLRW universe requires the consideration of a homogeneous and isotropic space-time described by the metric
\begin{equation}
    ds^2 = - N(t)^2dt^2 + a^2(t) \left(\frac{dr^2}{1-kr^2} + r^2 d\Omega_2^2 \right)\,.
\end{equation}
This fits naturally with the ADM formalism upon the choice for the shift functions:
\begin{equation}
    N^i = 0
\end{equation}
as well as the 3-dimensional metric $h$:
\begin{equation}
\begin{aligned}
    &h_{rr} = \frac{a^2 (t)}{1-kr^2}\\
    &h_{\theta \theta} = a^2(t)\,r^2\\
    &h_{\varphi \varphi} = a^2(t)\,r^2 \sin^2\theta\,.
\end{aligned}
\end{equation}
These are imposed by the Copernican Principle, i.e. homogeneity and isotropy. On the contrary, the lapse function $N$ is not determined a priori, so we will keep it free for now. It is related to the freedom in choosing the time coordinate. The square root of the 3-metric determinant is
\begin{equation}
    \sqrt{h} = \frac{a^3(t)\,r^2 \sin \theta}{\sqrt{1-kr^2}}\,.
\end{equation}
We can then compute the extrinsic curvature and find the conjugate momentum to the 3-metric:
\begin{equation}
    K^{ij} = \frac{1}{N} \frac{\dot{a}}{a} h^{ij} \quad \Pi^{ij} = - \frac{2}{N} \frac{\dot{a}}{a} \sqrt{h}h^{ij}
\end{equation}
and the corresponding traces:
\begin{equation}
	K = K^{ij}h_{ij} = \frac{3}{N} \frac{\dot{a}}{a}  \quad \Pi = \Pi^{ij}h_{ij} = -\frac{6}{N} \frac{\dot{a}}{a} \sqrt{h}\,,
\end{equation}
as well as the 3-dimensional Ricci scalar:
\begin{equation}
    ^{(3)}R = \frac{6k}{a^2}\,.
\end{equation}
The first Hamilton equation provides no additional information by itself, so the dynamics is obtained from the second Hamilton equation and the Hamiltonian constraint. Let us begin with the Hamiltonian constraint. We need the quantity:
\begin{equation}
	\Pi_{ij}\Pi^{ij} - \frac{1}{2}\Pi^2 = - \frac{6}{N^2} \left(\frac{\dot{a}}{a}\right)^2 h\,.
\end{equation}
Then:
\begin{equation}
	\mathcal{H} = - \sqrt{h}	  \frac{6k}{a^2} - \frac{6}{N^2}\sqrt{h} \left(\frac{\dot{a}}{a}\right)^2 = 2\kappa \frac{\delta \mathcal{L}_m}{\delta N}\,.
\end{equation}
The RHS is related with the stress-energy tensor in the following way:
\begin{equation}
	\frac{\delta \mathcal{L}_m}{\delta N} = -\frac{\delta \mathcal{L}_m}{\delta g_{00}} 2 N = - T^{00} N^2\sqrt{h}\,.
\end{equation}
Note the slightly different definitions for the contravariant and covariant stress-energy tensor, due to the sign flip of the functional derivative:
\begin{equation}
\begin{aligned}
	&T_{\mu \nu} = \frac{-2}{\sqrt{-g}} \frac{\delta \mathcal{L}_m}{\delta g^{\mu \nu}}\\
	&T^{\mu \nu} = \frac{2}{\sqrt{-g}} \frac{\delta \mathcal{L}_m}{\delta g_{\mu \nu}}\,.
\end{aligned}
\end{equation}
An FLRW universe is filled with a cosmological perfect fluid, whose stress-energy tensor is given by:
\begin{equation}
	T^{\mu\nu} = (\rho + p) u^{\mu} u^{\nu} + pg^{\mu \nu}\,,
\end{equation}
where the density $\rho$ and pressure $p$ are allowed to have a dependence on the entropy $S$ as well as on the scale factor $a$. Since the fluid that we are considering is isotropic and homogeneous, its 4-velocity is
\begin{equation}
	u^{\mu} = \left( \frac{1}{N},0,0,0\right) \quad g_{\mu \nu} u^\mu u^\nu =-1\,.
\end{equation}
Then we can identify the time-time component of the stress-energy tensor with the energy density of the fluid:
\begin{equation}
	 T^{00} = \rho N^{-2}	
\end{equation}
and the Hamiltonian constraint becomes
\begin{equation}
    \mathcal{H} = - \sqrt{h} \frac{6k}{a^2} - \sqrt{h} \frac{6}{N^2} \left(\frac{\dot{a}}{a}\right)^2 = -2\kappa \rho \sqrt{h}\,.
\end{equation}
This expression can be rearranged as
\begin{equation}
	\frac{1}{N^2} \left(\frac{\dot{a}}{a}\right)^2+\frac{k}{a^2} =\frac{8\pi G}{3} \rho
\end{equation}
This is the first Friedmann equation, which is nothing but a constraint on the dynamics of the FLRW space-time. If we make the choice $N=1$, which corresponds to the choice of cosmic time as time coordinate, we get the first Friedmann equation in its usual form:
\begin{equation}
    \left(\frac{\dot{a}}{a}\right)^2 + \frac{k}{a^2} = \frac{8\pi G}{3} \rho\,.
\end{equation}
Of course, one can work with conformal time and choose $N=a$. Then the first Friedmann equation becomes:
\begin{equation}
    \left(\frac{a'}{a}\right)^2 + k = \frac{8\pi G}{3} \rho a^2\,,
\end{equation}
which is consistent with a direct coordinate transformation.

Let us take a look now at the equation of motion involving the trace of the conjugate momentum:
\begin{equation}
\begin{aligned}
    \dot{\Pi} = & \dot{\Pi}^{ij}h_{ij} + \Pi^{ij}\dot{h}_{ij}\\
     = & - \frac{\delta \mathcal{H}_G}{\delta h_{ij}} h_{ij} + 2\kappa \frac{\delta \mathcal{L}_m}{\delta h_{ij}}h_{ij} - \kappa N \sqrt{h} \tilde{f}^{ij} h_{ij} + \Pi^{ij} \frac{\delta \mathcal{H}_G}{\delta \Pi^{ij}}\\
    = & \frac{1}{2} N \sqrt{h} ^{(3)}R + \frac{3}{2} N \sqrt{h} \left(K_{ij}K^{ij} - K^2\right)\\
    & + 2\kappa \frac{\delta \mathcal{L}_m}{\delta h_{ij}} h_{ij} - \kappa N \sqrt{h} \tilde{f}^{ij}h_{ij}\,.
\end{aligned}
\end{equation}
The only term left to compute in this expression is
\begin{equation}
K_{ij}K^{ij} - K^2 = - \frac{6}{N^2} \left(\frac{\dot{a}}{a}\right)^2\,.
\end{equation}
Then:
\begin{equation}
\begin{aligned}
	\dot{\Pi} = & \frac{1}{2}N\sqrt{h} \frac{6k}{a^2} + \frac{3}{2} N \sqrt{h} \left(-\frac{6}{N^2} \left(\frac{\dot{a}}{a}\right)^2\right)\\
	& + \kappa \sqrt{h}N T^{ij} h_{ij} - \kappa N \sqrt{h} \tilde{f}\,,
\end{aligned}
\end{equation}
where we used the definition of the spatial components of the stress-energy tensor:
\begin{equation}
	N \sqrt{h} T^{ij} = 2 \frac{\delta \mathcal{L}_m}{\delta h_{ij}}
\end{equation}
On the other hand, we have the geometric relation:
\begin{equation}
\begin{aligned}
	\dot{\Pi} & = - 2 \sqrt{h} \dot{K} - 2N\sqrt{h} K^2\\
	& = -6\sqrt{h} \left[ \frac{\ddot{a}}{aN} - \frac{\dot{a} \dot{N}}{aN^2} + 2 \left(\frac{\dot{a}}{a}\right)^2\right] \,,
\end{aligned}
\end{equation}
where we used
\begin{equation}
	\dot{K} = 3 \left(\frac{\ddot{a}}{aN} - \frac{\dot{a}^2}{a^2N} - \frac{\dot{a}\dot{N}}{aN^2}\right)\,.
\end{equation}
Using both expressions for $\dot{\Pi}$, the first Friedmann equation (i.e. the Hamiltonian constraint) and the spatial trace of the stress-energy tensor,
\begin{equation}
	\frac{1}{6} \kappa T^{ij}h_{ij} = \frac{4\pi G}{3} 3p\,,
\end{equation}
we get the equation of motion:
\begin{equation}
	\frac{\ddot{a}}{aN^2} - \frac{\dot{a}\dot{N}}{aN^3} = -\frac{4\pi G}{3} (\rho + 3p) + \frac{4\pi G}{3} N \tilde{f}\,.
\end{equation}
Of course, for $N=1$ this is nothing but the second Friedmann equation with an additional term of entropic origin. The effect of non-equilibrium thermodynamics is now encoded in the spatial trace:
\begin{equation}
    \tilde{f} = \tilde{f}^{ij}h_{ij}\,.
\end{equation}
This trace is related to the rate of entropy production by the second law of thermodynamics, encoded in the phenomenological constraint. Let us check how:
\begin{equation}
	\frac{1}{2\kappa}\frac{\delta L}{\delta S} \dot{S} = \frac{1}{2} a^3 \tilde{f}_{ij} \dot{h}^{ij} = -\dot{a} a^2 \tilde{f}\,.
\end{equation}
Introducing the temperature of the cosmological fluid,
\begin{equation}
		\frac{1}{2\kappa} \frac{\partial L}{\partial S} = -T\,,
\end{equation}
we get the expression for the trace of the entropic force:
\begin{equation}
	\tilde{f} = \frac{T\dot{S}}{\dot{a}a^2}\,.
\end{equation}
With this, the equation of motion becomes:
\begin{equation}
    \frac{\ddot{a}}{a} = -\frac{4\pi G}{3}( \rho + 3p ) + \frac{4\pi G}{3}\left(\frac{T\dot{S}}{a^2 \dot{a}}\right)\,,
\end{equation}
which is the second Friedmann equation modified by the enforcement of the second law of thermodynamics.

Most of the expansion history of the universe is adiabatic and thus remains unaffected by the inclusion of the effects of non-equilibrium thermodynamics in the Friedmann equations. Nevertheless, we can think of several phenomena in the expansion history during which entropy is copiously produced, such as the reheating of the universe, phase transitions and gravitational collapse to form black holes. We claim that these and other non-adiabatic phenomena in cosmology should be revisited, as their effect on the expansion rate may be non-negligible.

The assumption of homogeneity and isotropy implies that the tensor friction or entropic force $\tilde{f}_{ij}$ has only a trace component $\tilde{f}$. If we perturb around this solution, we expect the trace-less components to play a role as well, following the tensor decomposition in Eq.~(\ref{eq:ten_deco}). We leave the exploration of its consequences to a future publication where we will study the theory of cosmological perturbations in the presence of entropic forces arising from the trace, shear and vortical components of $f$.

Certain processes like gravitational collapse and structure formation are highly non-linear and cannot be understood within perturbation theory. It may be useful to treat this regime, highly non-linear and out of equilibrium, as an effective fluid. In this regard, we consider in the next section the similarities between generic entropic forces and the viscosity of a real fluid.

In principle, out of equilibrium phenomena could be incorporated into N-body simulations that study structure formation. This could be achieved by taking the non-relativistic limit of the non-equilibrium gravitational equations of motion in order to obtain a Newtonian plus entropic force.

\section{Real fluids in the variational formalism}

The results obtained in the previous sections are consistent with the relativistic dynamics of real fluids, i.e. fluids with viscosity and heat transfer. Such fluids are described by a stress-energy tensor that deviates from that of a perfect fluid \cite{LandauLifshitz2, LandauLifshitz6}
\begin{equation}
	T_{\mu \nu} = (\rho + p) u_\mu u_\nu + p g_{\mu \nu} + \tau_{\mu \nu}\,.
\end{equation}
This can be seen as a particular case of the variational formalism if the following equation is satisfied
\begin{equation}
	\tau_{\mu \nu} = - f_{\mu \nu}
\end{equation}
The additional term satisfies the orthogonality property
\begin{equation}
	u^\mu \tau_{\mu \nu} = 0\,.
\end{equation}
This is consistent with our description in the ADM formalism. In the comoving orthogonal gauge, $u^\mu = n^\mu$ and so the motion of the fluid is orthogonal to the constant time hypersurfaces.

For a vanishing chemical potential, the second law of thermodynamics of a real fluid takes the form
\begin{equation}
	T D_{\mu}(\sigma u^\mu) = \tau_\mu^\nu D_\nu u^\mu
\end{equation}
The LHS can be split into parallel and perpendicular components to the hypersurfaces, so that
\begin{equation}
\begin{aligned}
	D_{\mu}(\sigma u^\mu) & = n_\mu n^\alpha D_\alpha (\sigma u^\mu) + \nabla_\mu (\sigma u^\mu)\\
	& = n_\mu \pounds_n (\sigma u^\mu) + n_\mu \sigma u^\alpha D_\alpha n^\mu + \nabla_\mu (\sigma u^\mu) \,,
\end{aligned}
\end{equation}
where $\sigma$ is the local entropy density of the fluid. In the comoving orthogonal gauge this expression becomes
\begin{equation}
	D_{\mu}(\sigma u^\mu) = \pounds_n \sigma + \nabla_\mu (\sigma n^\mu)\,.
\end{equation}
On the other hand, in this gauge we have
\begin{equation}
	D_\mu u_\nu = \frac{1}{2} \pounds_n h_{\mu \nu}
\end{equation}
and the temperature of the fluid can be identified with
\begin{equation}
	T = - \frac{1}{N \sqrt{h}} \frac{\partial \mathcal{L}}{\partial s}
\end{equation}
We find then that the second law of thermodynamics of a real fluid can be rewritten in terms of the phenomenological constraint. This requires the following identifications
\begin{equation}
	f_{\mu\nu} = - \tau_{\mu\nu} \quad s^{tot} = \sigma \quad j_s^i = - \sigma u^i\,.
\end{equation}

We can still go one step further in the identification between the entropic force tensor and the viscosity tensor by inspecting its usual form
\begin{equation}
\begin{aligned}
	\tau_{\mu \nu} = & - \eta \left( D_\nu u_\mu + D_\mu u_\nu - u_\nu u^\alpha D_\alpha u_\mu - u_\mu u^\alpha D_\alpha u_\nu \right)\\
	& - \left( \zeta - \frac{2}{3} \eta \right) D_\alpha u^\alpha \left( g_{\mu \nu} + u_\mu u_\nu \right)\,,
\end{aligned}
\end{equation}
where $\eta$ and $\zeta$ are, respectively, the shear and bulk viscosity coefficients. Let us now focus on a homogeneous and isotropic fluid filling an FLRW universe. In this example, the covariant derivatives are given by
\begin{equation}
	D_\mu u^\nu = \frac{\dot{a}}{a} \delta_\mu^\nu\,,
\end{equation}
which means that the viscosity tensor is reduced to
\begin{equation}
	\tau_{\mu \nu} = - 3 \zeta \frac{\dot{a}}{a} h_{\mu \nu} \,.
\end{equation}
We can compare this with the expression of the trace of the entropic force obtained previously
\begin{equation}
	\frac{T\dot{S}}{a^3 H} = \tilde{f} = \tilde{f}^{ij}h_{ij} = - \tau_{ij} h^{ij}
\end{equation}
and obtain the following identity for the bulk viscosity coefficient
\begin{equation}
\zeta	= \frac{T\dot{S}}{9 H^2 a^3} > 0\,.
\end{equation}
Let us elaborate a bit on the results of this section. First of all, the conventional formulation of general relativistic real fluids can be recovered by means of the variational formulation of non-equilibrium thermodynamics in General Relativity. In fact, one does not even need to impose additional terms on the energy momentum tensor. Instead, they are effectively generated by simply assuming the pressure and the energy density of the fluid to have a dependency on the entropy.

The variational description allows the inclusion of dissipative effects to any matter or gravity content, as long as it has time-dependent entropy. This means that we can interpret non-equilibrium phenomena in General Relativity as an effective viscosity term of a real (i.e. non ideal) fluid. In this sense, our results allow for a variational, first principles formulation of real fluids and the generalization of their dissipative effects to arbitrary matter and gravity contents.

We point out that the variational and phenomenological constraints are imposed before obtaining the equations of motion and must be satisfied at all times. This is a fundamental difference with the theory of real fluids.

Here we considered a vanishing chemical potential, which means that we did not impose particle number conservation. This excludes thermal conduction effects. Nevertheless, one could in principle impose also particle number conservation at the Lagrangian \cite{GAYBALMAZ2017194}.

In the homogeneous and isotropic limit there is only bulk viscosity, parametrized by $\zeta$. However, shear viscosity, parametrized by $\eta$, may play a role in characterizing entropic forces in gravitational collapse and structure formation.

\section{Conclusions}

The final formulation of the theory of gravity in terms of the curvature of space and time took several decades to develop, at the beginning of last century, until the physical consequences of the theory were finally understood, and its limitations accepted, e.g. from black hole singularities to the origin of the universe. These developments brought forward new phenomena like gravitational waves and gravitational lensing which, together with gravitational redshift, allowed astronomers to map the universe. Nowadays, we have mastered both the theory and its observational consequences, and constructed a standard model of the macroscopic universe, which together with the standard model of particle physics, based on quantum field theory, provides a coherent picture over many orders of magnitude of time and energy.

However, in this global picture, the thermodynamical notions of temperature and entropy appeared always in the context of thermal equilibrium and adiabatic expansion. Far-from-equilibrium phenomena, like the gravitational collapse of matter structures or the reheating of the universe after inflation, were treated discontinuously, and were assumed not to modify the local space-time structure.

In this first work, we have developed a generally-covariant formulation of out-of-equilibrium thermodynamics in General Relativity. For this we have introduced thermodynamics as a constraint on the Lagrangian density and derived the coupled differential equations via a variational principle. The generalization to curved manifolds was straightforward and thus we derived the modified Einstein field equations, which presented an extra term together with the matter content that takes into account the non-equilibrium dynamics. In particular, the Bianchi identities imply the covariant non-conservation of the energy momentum tensor, reflecting the presence of covariant entropic forces associated with the non-equilibrium dynamics.

In order to explore better the physical consequences of these entropic forces we have developed a Hamiltonian formulation of the dynamics {\it a la} ADM, by making a (3+1)-splitting of space-time. This has allowed us to show that the Hamiltonian constraint remains unchanged and the entropic forces appear only in the dynamical equations of motion.

We have also derived the corresponding Raychaudhuri equations for a congruence of geodesics and found that for matter satisfying the strong energy condition, it is possible that an expanding congruence avoids recollapse for a sufficiently large and positive entropic contribution. This could have important consequences for the formation of black holes and the expansion of the universe.

Furthermore, we have realised that there can be not only ``bulk" entropic contributions associated with hydrodynamical matter, but also ``boundary" contributions associated with surface terms like those of black hole horizons, which contribute also to the energy density.

There is still a lot of work to do. All of those out-of-equilibrium situations where entropy is produced, like in the gravitational collapse of matter, or the copious particle production in the early universe, may induce large entropic forces responsible for local or global accelerations, which may then modify the structure of space-time. In particular, we explore in paper II the effect that cosmological horizons have on the present cosmic acceleration, and in paper III the secondary burst of accelerated expansion at preheating after inflation.

The covariant formulation of entropic forces opens the exploration of all relevant out-of-equilibrium processes in the universe and will surely generate completely new and unexpected phenomenology. We are looking forward to these developments.

\begin{acknowledgements}

The authors acknowledge support from the Spanish Research Project PGC2018-094773-B-C32 (MINECO-FEDER) and the Centro de Excelencia Severo Ochoa Program SEV-2016-0597. The work of LEP is funded by a fellowship from ``La Caixa" Foundation (ID 100010434) with fellowship code LCF/BQ/IN18/11660041 and the European Union Horizon 2020 research and innovation programme under the Marie Sklodowska-Curie grant agreement No. 713673.
\end{acknowledgements}

\appendix*

\section{Variational formulation of entropic forces in the continuum}

In this appendix we connect the variational formulation for continuous systems described by Gay-Balmaz and Yoshimura \cite{GAYBALMAZ2017194} with the shortcut used in the Hamiltonian formulation of non-equilibrium thermodynamics in General Relativity.

The original rigorous variational formulation of non-equilibrium thermodynamics requires the addition of a new term in the Lagrangian and the introduction of new variables
\begin{equation}
	\delta \int d^4x \sqrt{-g} \left( \mathcal{L} + (S-\Sigma) \dot{\Gamma} \right) = 0\,.
\end{equation}
$\Gamma$ is the thermodynamic displacement, while $S$ and $\Sigma$ are functions whose time derivatives will be linked to internal and total entropy production. The variation of the action is equal to
\begin{equation}
	\int d^4x \sqrt{-g} \left(\frac{\delta \mathcal{L}}{\delta \phi}\delta \phi + \frac{\partial \mathcal{L}}{\partial S} \delta S - (\dot{S} - \dot{\Sigma})\delta \Gamma + (\delta S - \delta \Sigma) \dot{\Gamma} \right)\,.
\end{equation}
The stationary-action principle is now supplemented with the phenomenological and variational constraints
\begin{equation}
\begin{aligned}
	&\frac{\partial \mathcal{L}}{\partial S} \dot{\Sigma} = - P^{i} \nabla_{i} \dot{\phi} + J^i\nabla_i \dot{\Gamma}\\
	&\frac{\partial \mathcal{L}}{\partial S} \delta\Sigma = - P^{i} \nabla_{i} \delta \phi + J^i\nabla_i \delta\Gamma\,,\\
\end{aligned}
\end{equation}
which assume that there is no external power supply. We will restrict ourselves to the simple case where the field is scalar and so the tensor $P$ is a vector, but it could have a higher order. In the original work the authors describe a fluid in its material representation. In that case $\phi$ is a position vector and $P$ is a 2-tensor.

With this constraints, the variation of the action is equal to
\begin{equation}
\begin{aligned}
	\int d^4x \sqrt{-g} &\bigg[ \left(\frac{\delta \mathcal{L}}{\delta \phi} + \dot{\Gamma} \left(\frac{\partial \mathcal{L}}{\partial S}\right)^{-1} P^i \nabla_i\right) \delta \phi\\
	& +\left( \dot{\Gamma} + \frac{\partial \mathcal{L}}{\partial S} \right)\delta S\\
	& + \left( \dot{\Sigma} - \dot{S} - \dot{\Gamma} \left(\frac{\partial \mathcal{L}}{\partial S}\right)^{-1} J^i \nabla_i \right)\delta \Gamma \bigg] = 0\,.
\end{aligned}
\end{equation}
Integrating by parts we get
\begin{equation}
\begin{aligned}
	\int d^4x \sqrt{-g} &\bigg[ \left(\frac{\delta \mathcal{L}}{\delta \phi} - \nabla_i \left( \dot{\Gamma} \left(\frac{\partial \mathcal{L}}{\partial S}\right)^{-1} P^i \right) \right) \delta \phi\\
	&+\left( \dot{\Gamma} + \frac{\partial \mathcal{L}}{\partial S} \right)\delta S\\
	& + \left( \dot{\Sigma} - \dot{S} + \nabla_i \left( \dot{\Gamma} \left(\frac{\partial \mathcal{L}}{\partial S}\right)^{-1} J^i\right) \right)\delta \Gamma \bigg] = 0\,.
\end{aligned}
\end{equation}
The equations of motion are obtained as usual by noticing that each variation is independent. From $\delta S$ we get
\begin{equation}
	\dot{\Gamma} = - \frac{\partial \mathcal{L}}{\partial S} \equiv T\,,
\end{equation}
while $\delta \phi$ and $\delta \Gamma$ give
\begin{equation}
\begin{aligned}
	\frac{\delta \mathcal{L}}{\delta \phi} + \nabla_i P^i = 0\\
	\dot{\Sigma} = \dot{S} + \nabla_i J^i\,.
\end{aligned}
\end{equation}
These equations of motion and the phenomenological constraint fully describe the time evolution of the system. The last one is the entropy balance equation. Even though we imposed it earlier, this shows that it can also be derived from the stationary-action principle.

We will now perform a variable transformation that will affect only the phenomenological constraint, in order for it to have the form used throughout the paper. Note that the variational condition is invariant under the following redefinitions
\begin{equation}
\begin{aligned}
	\frac{\partial\mathcal{L}}{\partial S} \delta \Sigma \rightarrow \frac{\partial\mathcal{L}}{\partial \Sigma} \delta \Sigma + \nabla_i \left(J^i \delta \Gamma \right) - \nabla_i \left(P^i\dot{\phi}\right) \\
	\frac{\partial\mathcal{L}}{\partial S} \delta S \rightarrow \frac{\partial\mathcal{L}}{\partial S} \delta S + \nabla_i \left(J^i \delta \Gamma \right) - \nabla_i \left(P^i\dot{\phi}\right) \\
\end{aligned}
\end{equation}
and the equivalent changes in the derivatives
\begin{equation}
\begin{aligned}
	&\frac{\partial\mathcal{L}}{\partial S} \dot{S} \rightarrow \frac{\partial\mathcal{L}}{\partial S} \dot{S} + \nabla_i \left(J^i \dot{\Gamma} \right) - \nabla_i \left(P^i\dot{\phi}\right)\\
	&\frac{\partial\mathcal{L}}{\partial S} \dot{\Sigma} \rightarrow \frac{\partial\mathcal{L}}{\partial S} \dot{\Sigma} + \nabla_i \left(J^i \dot{\Gamma} \right) - \nabla_i \left(P^i\dot{\phi}\right)\,.\\
\end{aligned}
\end{equation}
These replacements add total derivatives to the varied Lagrangian and so have no physical effect. The equations of motion stay the same. In fact, these are the kind of terms added when integrating by parts before obtaining the equations of motion. The temperature also stays the same, since $\partial S / \partial S' = 1$, being $S'$ the newly defined entropy.

As already mentioned, the only final equation that is transformed is the phenomenological constraint, which becomes
\begin{equation}
	T \dot{\Sigma} = \nabla_i P^i \dot{\phi} + T \nabla_i J^i\,,
\end{equation}
that is
\begin{equation}
	T \dot{S} = \nabla_i P^i \dot{\phi}\,,
\end{equation}
where $\nabla_i P^i$ is what we called the friction or entropic force tensor, only that it is a scalar here. In our formulation we replace the time derivative by a more covariant notion of time evolution, the Lie derivative along the normal vector~$n$.

We conclude that both formulations are equivalent. Our choice is motivated by simplicity and physical intuition, as a homogeneous and isotropic entropy function becomes a particular case in a clearer way.

\bibliographystyle{h-physrev}
\bibliography{paperEF}

\end{document}